\documentclass[
reprint,
nofootinbib,
amsmath,amssymb,
aps,
prx,
twocolumn,
floatfix,
]{revtex4-2}

\usepackage{algorithm}
\usepackage{amssymb}   
\usepackage{amsfonts}
\usepackage{amsmath}
\usepackage{mathtools}
\usepackage{MnSymbol}
\usepackage{dsfont}
\usepackage{dcolumn}   
\usepackage{bbm}
\usepackage{bm}        
\usepackage[mathscr]{eucal}
\usepackage{placeins}

\usepackage{algorithmic}
\usepackage{braket}
\usepackage{subcaption}
\usepackage{graphicx}
\usepackage[usenames,dvipsnames]{xcolor}

\usepackage{verbatim}

\usepackage{parskip} 

\usepackage{tikz}
\usetikzlibrary{quantikz}

\usepackage{hyperref}
\definecolor{bblue}{RGB}{0, 126, 204}
\definecolor{dgreen}{RGB}{0,102,51}
\definecolor{blueus}{RGB}{10, 49, 97}
\definecolor{teal}{RGB}{0,128,128}
\definecolor{orangered}{RGB}{255,69,0}
\hypersetup{
	unicode=false,          
	pdftoolbar=true,        
	pdfmenubar=true,        
	pdffitwindow=false,     
	pdfstartview={FitH},    
	pdftitle={qalgorithm},    
	pdfauthor={},     
	pdfsubject={},   
	pdfcreator={},   
	pdfproducer={}, 
	pdfkeywords={} {} {}, 
	pdfnewwindow=true,      
	colorlinks=true,       
	linkcolor=purple, 
	citecolor=bblue,        
	filecolor=magenta,      
	urlcolor=teal
}

\graphicspath{ {./figures/} }

\tikzset{
operator/.append style={fill=dgreen!20},
my label/.append style={above right,xshift=0.3cm},
phase label/.append style={label position=above}
}

\DeclareMathOperator*{\argmin}{argmin}

\begin{document}

\preprint{APS/123-QED}

\title{Quantum Markov chain Monte Carlo method with programmable quantum simulators
}%

\author{Mauro D'Arcangelo}
\email{mauro.darcangelo@pasqal.com}
\author{Younes Javanmard}
\author{Natalie Pearson}
\affiliation{Pasqal, 24 rue Emile Baudot, 91120 Palaiseau, France}%

\date{\today}

\begin{abstract}
In this work, we present a quantum Markov chain algorithm for many-body systems that utilizes a special phase of matter known as the Many-Body Localized (MBL) phase. We show how the properties of the MBL phase enable one to address the conditions for ergodicity and sampling from distributions of quantum states. We demonstrate how to exploit the thermalized-to-localized transition to tune the acceptance rate of the Markov chain, and apply the algorithm to solve a range of combinatorial optimization problems of quadratic order and higher. The algorithm can be implemented on any quantum processing unit capable of simulating the Floquet dynamics of a one-dimensional Ising chain with nearest-neighbor interactions, providing a practical way of sampling from thermal distributions of Hamiltonians that cannot be natively implemented on the quantum hardware.

\end{abstract}

\maketitle


\section{Introduction}

Near-term, so-called NISQ (Noisy Intermediate Scale Quantum) devices are limited by scale and errors, and it has proven challenging to find useful applications for them. The most promising evidence of an advantage over classical algorithms is in quantum simulation, where the analog operation of the devices has enabled the simulation of quantum systems beyond the ability of classical computers \cite{Mostame2016, Johnson2011, King2024, Semeghini2021}, with the caveat that linking achievable simulations with the solving of real-world problems is not trivial. In general, computation on a quantum computer can be described as sampling; it is necessary to calculate the values of observables in quantum simulation, to measure the result of a combinatorial optimization or factorization problem, and is generally equivalent to a search problem \cite{aaronson2014equivalence}.
\begin{figure}[h!]
    \includegraphics[width=\columnwidth]{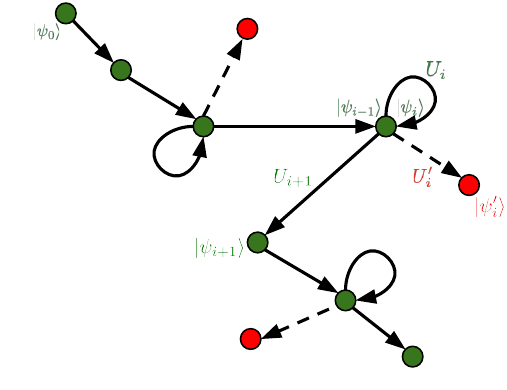}
    \label{mc moves}
  \caption{
The figure illustrates the schematics of the algorithm. The algorithm iteratively proposes new quantum states starting from an initial state $\ket{\psi_0}$. At the $i$-th iteration, a new state $\ket{\psi_i'}$ is proposed by evolving $\ket{\psi_{i-1}}$ with a unitary $U_i'$. The move is evaluated in a Metropolis accept/reject scheme. If accepted $U_i=U_i'$ and $\ket{\psi_i}=\ket{\psi_i'}$, if rejected $U_i=I$ and $\ket{\psi_i}=\ket{\psi_{i-1}}$. Green dots and solid arrows represent accepted moves, while red dots and dashed arrows represent rejected moves. Each move is generated by a unitary operator in the many-body localized (MBL) phase, which allows a controlled exploration of the Hilbert space.
}
\label{fig1: mcmc-mbl alg}
\end{figure}
A significant challenge of analog quantum computation is that, although theoretically universal, it is usually restricted to a small class of Hamiltonians realizable by particular hardware. In this work, we introduce a \emph{novel Quantum Markov Chain algorithm} that dramatically extends the range of simulable Hamiltonians and, consequently, the range of problems solvable using realistic analog quantum hardware. Our algorithm, illustrated in Fig.~\ref{fig1: mcmc-mbl alg}, consists of implementing a Markov chain in the Hilbert space of a qubit system. The algorithm allows to extract a chain of quantum states, or \emph{samples} in Markov chain terminology (not to be confused with samples as in measurements of a quantum state), each obtained from the previous one in an iterative fashion, up to a certain maximum chain length (i.e. number of iterations, or equivalently number of samples). Markov chains are widely used, often as a tool in Monte Carlo sampling, to explore a probability landscape by making a series of small moves that are accepted or rejected based on their probability, resulting in convergence to the target probability distribution. In our algorithm, these moves are performed by evolving a quantum system in a Hilbert space and then measuring its state to determine the probability of the move classically. This probability is then used to decide whether to accept or reject the move.

Markov chain Monte Carlo requires ergodic exploration, which is difficult to replicate in quantum dynamics. The main development introduced in this work overcomes this challenge by building upon recent approaches that exploit many-body localization (MBL) in the context of Hamiltonian engineering \cite{mbl_bayesian, Bastidas2022}. These works demonstrate how suitable external driving and system design can confine dynamics to the MBL regime. We use the stability of the MBL phase to construct localized transitions, enabling controlled sampling over time via an ergodic Markov chain of distinct non-ergodic MBL unitaries.

Many-body localization is a phenomenon that prevents an isolated quantum system from reaching thermal equilibrium. In general, isolated quantum systems are expected to relax to a thermal state at the level of each individual eigenstate, according to the eigenstate thermalization hypothesis (ETH)~\cite{Rigol2008}. However, disorder can induce significant deviations from conventional ergodic behavior through the phenomenon of many-body localization \cite{Anderson1958,Basko2006,Oganesyan2007,Nandkishore2015,Abanin2017}. In the MBL phase, strong disorder and interactions prevent thermalization, causing eigenstates to violate the ETH and allowing local observables to retain memory of initial conditions indefinitely. Moreover, MBL dynamics is marked by slow, logarithmic growth of entanglement entropy and can be described by an extensive set of local integrals of motion \cite{Serbyn2013,Huse2014,Imbrie2016review}.

By evolving the system under MBL unitaries and accepting or rejecting these steps according to the traditional Metropolis criterion \cite{metropolis, hastings}, we have realized a hybrid algorithm that traverses state space to sample from a desired probability distribution and serve as a way for optimal conversion from classical to quantum randomness~\cite{randomness}. We stress that the quantum dynamics induced by MBL is by definition non-ergodic in the physical sense. However, we exploit the properties of MBL in conjunction with Markov chain theory to artificially construct ergodicity in the algorithmic sense. We demonstrate the effectiveness of this algorithm in solving several mathematical problems of interest, corresponding to the preparation of ground states of nonnative Hamiltonians with high probability.

The paper is organized as follows. First, we review the relevant prior art on existing quantum Markov chain formulations. We then introduce the methodology used, including the implementation of MBL unitaries on realistic quantum hardware through Floquet engineering. This is followed by the results we obtain using this method in combinatorial optimization to solve the Max-Cut, Maximum Independent Set (MIS), and integer factorization problems. Finally, we discuss the challenges and implications of this algorithm and suggest potential directions for future work.

\section{Quantum Markov chains}\label{sec:priorart}

A Markov chain is a stochastic process to traverse a state space where the probability of exploring a certain state depends only on the state explored in the previous step \cite{Norris1997, Krauth2006}. Markov chains have been extensively proven to be a remarkably flexible tool, serving as the backbone of various Monte Carlo sampling techniques. The original applications were mainly focused on statistical physics \cite{Newman1999} and high-energy physics \cite{munster}, but have since paved the way for a plethora of successful Monte Carlo algorithms~\cite{langevin, hmc, parallel_tempering_1, parallel_tempering_2} that nowadays extend to other fields such as hyperparameter tuning in Bayesian inference \cite{nuts} and combinatorial optimization through simulated annealing \cite{kirkpatrick, Martonak}.

Given the success and widespread use of Markov chain Monte Carlo methods, it is no surprise that several quantum versions of it currently exist, some claiming advantage in the form of a Grover-type quadratic speed-up for quantum walks \cite{szegedy}. A recent comprehensive account of quantum Markov chain methods is given in Ref.~\cite{walking_review}.

The first attempt to formulate a quantum Markov chain was a quantum Metropolis algorithm \cite{temme} aimed at performing an ergodic walk on the eigenstates of a quantum Hamiltonian. However, it made use of a boosted and shift-invariant version of quantum phase estimation that was later proven to be impossible, as shown in Appendix H of~\cite{CKBG}. In addition, implementing the rejection part of Metropolis in a quantum setting is especially complicated due to the no-cloning theorem, and circumventing it with the Marriott-Watrous rewinding algorithm \cite{MW} complicates the implementation considerably. Quantum Metropolis was made more robust in a recent proposal~\cite{jiang2024}, where the dependency on the Marriott-Watrous rewinding algorithm is eliminated, and a non-shift-invariant form of quantum phase estimation is used. Although this later proposal is theoretically sound, near-term implementability remains in question because of the heavy use of quantum phase estimation and the unfavorable scaling with problem size and precision.

These works focused on providing a viable quantum version of Metropolis capable of targeting, among other things, Gibbs distributions of the form $e^{-\beta H}$, where $H$ is a specific Hamiltonian and $\beta$ is the inverse temperature. An alternative direction for quantum Markov chains is to directly replicate the physical thermalization process of a quantum system via the Lindbladian evolution of a system in contact with a bath, formally related to a continuous-time Markov chain, where the Lindbladian naturally handles the rejection step. These efforts culminated in a rigorous formulation with provable quantum detailed balance guarantees for noncommuting Hamiltonians \cite{CKG, CKBG} and a discrete-time version of it \cite{discreteCKG}. These works are mostly theoretical in nature, and near-term feasibility is unclear. Related work~\cite{movassagh}, similarly based on a bath thermalization process, provides a concrete implementation on current-generation quantum hardware but requires a QPU capable of simulating the desired Hamiltonian.

In general, the methods described so far require hardware resources and precision that are not currently attainable for problems sizes of practical interest. In an effort to provide algorithms of more immediate applicability, some authors have developed workflows where classical and quantum resources are combined in a hybrid quantum-classical Markov chain. 

An example of this is a recent proposal where a classical Markov chain is combined with a quantum update to generate configurations of classical spin Hamiltonians~\cite{Layden2023}. This method has the benefit of not requiring long coherence times of the quantum hardware, since quantum evolutions are only used to generate a single move of the Markov chain, and the accept/reject step is easily calculated classically. However, once again, the quantum hardware is required to implement the desired Hamiltonian. Since then, a fully classical algorithm based on \cite{Layden2023} has been developed \cite{inspired}. In a closely related development, the quantum-enhanced algorithm has been extended to variational Monte Carlo~\cite{qavmc}.

Another recent work in this hybrid direction~\cite{gradient_mcmc} proposes a quantum method to improve the classical Langevin and Hamiltonian Monte Carlo. They identified gradient calculation as the bottleneck of the algorithm and proposed using a quantum oracle to obtain an estimate of the gradient. 

Our Quantum Markov chain proposal targets distributions of quantum states through Metropolis, and in this regard is similar to~\cite{temme}. However, to avoid a quantum phase estimation step which would make the algorithm difficult to implement, we resorted to a hybrid solution where the accept/reject step is delegated to classical processing. It is important to remark that other hybrid solutions, such as~\cite{Layden2023}, in contrast with our work, target classical spin configurations, and the QPU is used as a sampler from which only one bitstring at a time is needed.  In our case instead, we exploit information from the full quantum state, which requires more extensive sampling of the quantum state at each step. In addition, state preparation necessarily involves evolving the initial state through the whole chain every time. These are the main challenges posed by our algorithm, stemming from the choice of trading off algorithmic complexity for sampling complexity.

\section{Algorithm and the Hamiltonian Set-Up}

Assume that one wants to sample $N$-qubit quantum states $\ket{\psi}$ according to some distribution $D(\ket{\psi})$, defined as the expectation value $\bra{\psi}O\ket{\psi}$ of some positive-definite observable $O$. The typical example is the Gibbs distribution $D(
\ket{\psi}) \sim \bra{\psi}e^{-\beta H}\ket{\psi}$ for some Hamiltonian $H$ and inverse temperature $\beta>0$. We propose a Markov chain where the moves are generated by quantum evolutions in the Hilbert space and are evaluated in a classical Metropolis accept/reject scheme. The algorithm works as follows. First, a register of $N$ qubits is initialized in an arbitrary initial state $\ket{\psi_0}$, typically $\ket{\psi_0} = \ket{0}^{\otimes N}$. A new state $\ket{\psi_1'}$ is obtained by applying a random unitary $U_1'$ to $\ket{\psi_0}$. The move is evaluated by calculating the Metropolis ratio $D(\ket{\psi_1'})/D(\ket{\psi_0})$. If the move is accepted, we set $\ket{\psi_1} = \ket{\psi_1'}$, otherwise $\ket{\psi_1} = \ket{\psi_0}$. The steps are then repeated until the desired chain length is reached.

It is important to mention that on quantum hardware at step $i$ it is not possible to apply directly the unitary $U_i'$ to the previous state, since state $\ket{\psi_{i-1}}$ would have been destroyed during measurement. Therefore, at each step the whole chain of unitaries needs to be reapplied to the initial state $\ket{\psi_0}$.

The original contribution of this work was to utilize many-body localization as an effective method to propose moves. To explore the Hilbert space ergodically, the proposed moves must define an irreducible and reversible Markov chain. Although reversibility is not strictly necessary, it is a straightforward way to satisfy detailed balance when combined with the Metropolis accept/reject rule. In practice, it is also important to be able to tune the ``length'' of the moves to enforce an acceptance rate (the ratio between the number of accepted moves and the number of proposed moves) that minimizes the mixing time of the chain \cite{mixing_time_1, mixing_time_2, mixing_time_3, mixing_time_4}. In the following, we explain how the dynamics of many-body localization is instrumental in producing moves in the Hilbert space that meet these technical requirements.

The irreducibility of a chain of randomly sampled distinct MBL unitaries follows from a random matrix theory argument. The thermalized and many-body localized phases—and the transition between them—are well described by the random matrix theory of power law banded matrices (PRBM) \cite{prbm}. In random matrix theory, phases are determined by the level spacing statistics of the ensemble: for example, unitaries in the thermalized phase display the level spacing statistics typical of the Circular Orthogonal Ensemble (COE), while MBL unitaries display Poissonian statistics. The critical observation that we leverage to prove irreducibility is that products of many matrices in the Poissonian (MBL) phase exhibit the level spacing statistics of the Circular Unitary Ensemble (CUE). CUE represents the uniform measure for unitary operators~\cite{quantum_signatures_chaos}, which is the ensemble where any unitary operator is equiprobable. Therefore we conclude that a Markov chain of MBL unitaries can in principle represent any unitary operator in the Hilbert space, making it irreducible. See Appendix~\ref{app:theory} for a formal proof of irreducibility and an analysis on convergence bounds of the chain, and Appendix~\ref{app:cue} for details on the speed of convergence of products of MBL to CUE.

In principle, since products of MBL unitaries converge to CUE, one might ask why not use unitaries in the thermalized phase in the first place, which converge to CUE even faster. The reason is a practical one. Proposing new states with thermalization unitaries amounts to making uniformly random guesses in an exponentially large Hilbert space. MBL, on the other hand, allows moving slowly but surely towards the relevant sector of the Hilbert space, making it a more effective tool for Hilbert space exploration. A similar idea is discussed in \cite{mbl_bayesian} from the perspective of machine learning trainability, where it is shown that optimizing over the MBL parameter space allows to learn much more effectively than doing the same in the thermalized phase.

While MBL was initially studied in time-independent systems, it also persists in periodically driven (Floquet) settings, leading to \emph{Floquet-MBL} phases \cite{Ponte2015,Lazarides2015,Abanin2016}. This allows implementation on realistic quantum hardware such as neutral atom quantum processors, enabling reversibility and control of move length in a practical setting.

Following \cite{PhysRevB.103.165132}, we consider the Floquet transverse-field Ising model, described by the Hamiltonian
\begin{align}
    H(t) = \sum_{i=1}^N h_i Z_i + \sum_{i=1}^N B(t) X_i + J \sum_{i=1}^{N-1} Z_i Z_{i+1}, 
    \label{Eq.1}
\end{align}
where $X_i$ and $Z_i$ are Pauli operators on site $i$ of a chain with $N$ sites. The interaction strength $J$ governs nearest-neighbor interactions, while the time-dependent field $B(t) = B_0 + \delta B \cos(\omega t)$ oscillates with frequency $\omega = 2\pi/T$. The disordered static fields $h_i$ are random variables uniformly sampled from the interval $[-W/2, W/2]$, where $W$ is known as the 
\emph{disorder strength}.

\begin{algorithm}[H]
\caption{Markov chain MBL Algorithm}
\begin{algorithmic}[1]\label{alg: mcmcmbl}
    \STATE \textbf{objective:} Take $M$ samples of $N$-qubit quantum states from a distribution $D(\ket{\psi})\propto\bra{\psi}O\ket{\psi}$ for some positive-definite observable $O$.
    \item[]
    \STATE \textbf{initialize:} Set the initial state of the system $\ket{\psi_0}$ and set parameters $W$, $J$, $B_0$, $\delta B$ and $\omega$ of Eq.~\eqref{Eq.1} to be in MBL phase.
    \item[]
    \FOR{$i=1$ \TO $M$}
        \STATE \textbf{draw a Hamiltonian:} For each qubit $j=1,\ldots,N$, select coefficients $h_j$ from a uniform distribution over $[-\frac{W}{2}, \frac{W}{2}]$.
        \STATE \textbf{time evolution:} Evolve the system using the Hamiltonian defined in Eq.~\eqref{Eq.1} using the chosen coefficients $h_j$:
        \[
        \ket{\psi_i'} = U_i' \ket{\psi_{i-1}} = U_i' U_{i-1} \ldots U_1\ket{\psi_0},
        \]
        where each $U$ is given by Eq.~\eqref{eq: unitary}. This step is implemented on a QPU.
        \STATE \textbf{measurement:} Calculate the expectation value $\bra{\psi_i'}O\ket{\psi_i'}$ from $K$ repeated measurements.
        \STATE \textbf{decision:} 
        \begin{ALC@g}
            \STATE Calculate acceptance probability $q$:
            \[q := \frac{D(\ket{\psi_i'})}{D(\ket{\psi_{i-1}})} = \frac{\bra{\psi_i'}O\ket{\psi_i'}}{\bra{\psi_{i-1}}O\ket{\psi_{i-1}}}\]
            \IF{$q\ge1$}
                \STATE Accept the current state transition: $\ket{\psi_i} = \ket{\psi_i'}$ and $U_i = U_i'$.
            \ELSE
                \STATE Draw a random number $r$ uniformly in $[0, 1]$.
                \IF{$r < q$}
                    \STATE Accept the current state transition: $\ket{\psi_i} = \ket{\psi_i'}$ and $U_i = U_i'$.
                \ELSE
                    \STATE Reject the current state transition: $\ket{\psi_i} = \ket{\psi_{i-1}}$ and $U_i = \mathbb{I}$.
                \ENDIF
            \ENDIF
        \end{ALC@g}
    \ENDFOR
\end{algorithmic}
\end{algorithm}

The Floquet unitary captures the stroboscopic dynamics over one driving period
\begin{align}\label{eq: unitary}
    U = \mathcal{T} \exp\left[-i \int_0^T H(t) dt\right] \equiv \exp\left[-i H_F T\right],
\end{align}
where $\mathcal{T}$ denotes time-ordering and $H_F$ is the effective Floquet Hamiltonian. Reversibility is ensured by the fact that the inverse dynamics is easily obtained by inverting every sign in the Hamiltonian. Therefore, if it is possible to reverse the sign of the transverse field and interaction terms, and the random fields $h_i$ are sampled evenly from a symmetric interval, the Markov chain is reversible.

Using MBL unitaries allows us to tune the acceptance rate of the Markov chain. Increasing the disorder strength $W$, the system undergoes a phase transition from the thermalized phase to the MBL phase~\cite{PhysRevB.103.165132}. Thermalization and localization are opposite in terms of Hilbert space exploration: a system evolving under a unitary in the thermalized regime will end up in a uniformly random state of the Hilbert space, while a system evolving under MBL will retain memory of the initial conditions. For this reason, we argue that the disorder strength $W$ controls the average ``length'' of the move defined by the unitary, and therefore the acceptance rate of the Markov chain. This statement is investigated numerically in the Results section.

See Algorithm~\ref{alg: mcmcmbl} for a breakdown of each step.

We stress that the algorithm is suitable for both analog and digital quantum devices. However, it is natively written as Hamiltonian evolution, and thus its potential for near-term applications relies mainly on the analog version. In addition, analog NISQ algorithms have been shown to be more resilient to errors compared with their Trotterized digital counterparts~\cite{trotterization, emulating_annealing, proliferation}. The results presented in the next section refer to the analog version of the algorithm.

\section{Results}\label{sec:results}
As a use case, we consider sampling from a distribution determined by the cost function of optimization problems in the form of a general $k$-body Hamiltonian:
\begin{align}
    H = &\sum_i J_i Z_i + \sum_{i<j} J_{ij} Z_i Z_j + \sum_{i<j<k} J_{ijk} Z_i Z_j Z_k \nonumber
\\ &+ \sum_{i<j<k<l} J_{ijkl} Z_i Z_j Z_k Z_l + \dots, 
\label{Eq: Optimization problems}
\end{align}
where $Z_i$ are the Pauli-$Z$ operators.

The popular use of NISQ analog devices as programmable quantum annealers generally limits them to solving problems mapped onto quadratic Hamiltonians~\cite{hubo_eig}. In our work, we use the dynamics of many-body localization to set up a Markov chain capable of sampling directly from the low-temperature regime of the k-body Hamiltonian of Eq.\eqref{Eq: Optimization problems}, without requiring mappings from k-body to 2-body interactions. 

We focus first on two well-known graph combinatorial optimization problems: Maximum Independent Set (MIS) and Max-Cut~\cite{BOROS} (see Appendix \ref{sec: mis} and \ref{sec: maxcut} for the associated cost functions). We study these problems on Erd\H{o}s-R\'enyi graphs with an edge probability of 0.5. These graphs are a standard model for studying random graph properties and phase transitions in combinatorial optimization and represent hard instances for both MIS and Max-Cut~\cite{er_hardness}. To showcase the ability to go beyond quadratic Hamiltonians, we then turn to a combinatorial optimization formulation of the prime factorization problem~\cite{factorization}, which contains up to 4-body interaction terms (see Appendix \ref{sec: fact}).

The strategy we use to find the solution to a combinatorial optimization problem encoded in a Hamiltonian of the type \eqref{Eq: Optimization problems} is to sample from its Gibbs distribution $e^{-\beta H}$ for large $\beta$ (low temperature), that is peaked around the ground state of the Hamiltonian.

The MBL unitaries are generated by taking one period of the Hamiltonian~\eqref{Eq.1}, with $B_0 = -\delta B = 1.25 J$ and $\omega = 10J$. Nearest-neighbour interaction is set to $J = 4.15$, and the initial state is chosen to be $\ket{\psi_0}=\ket{0}^{\otimes N}$. For a discussion on parameter selection criteria, see Appendix \ref{app:parameters}.

\subsection{Thermalization}

\begin{figure}[htpb]
\centering
\begin{subfigure}{.5\textwidth}
    \caption{}
    \includegraphics[width=\linewidth]{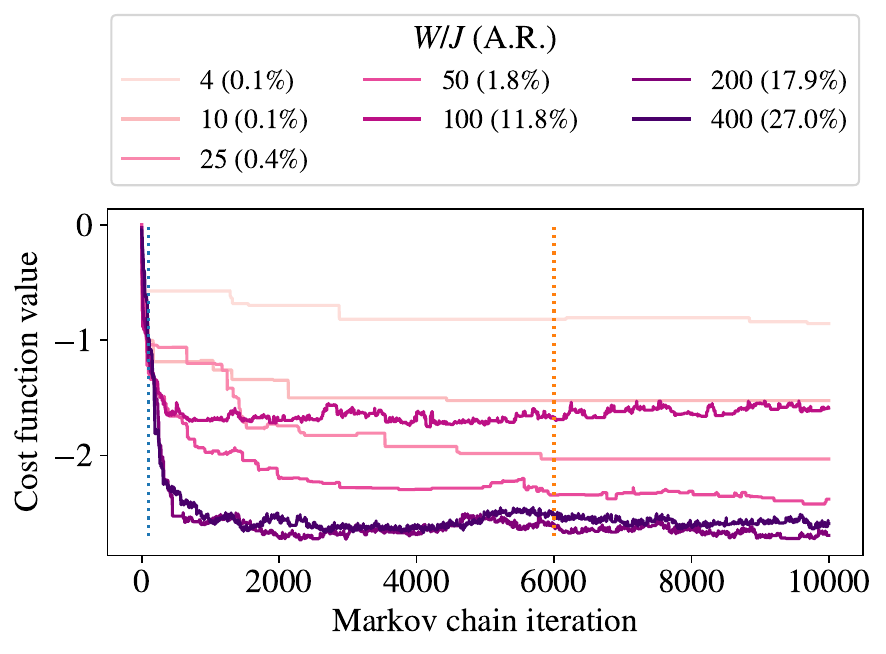}
    \label{fig:thermalization}
\end{subfigure}
\begin{subfigure}{.5\textwidth}
    \caption{}
    \includegraphics[width=\linewidth]{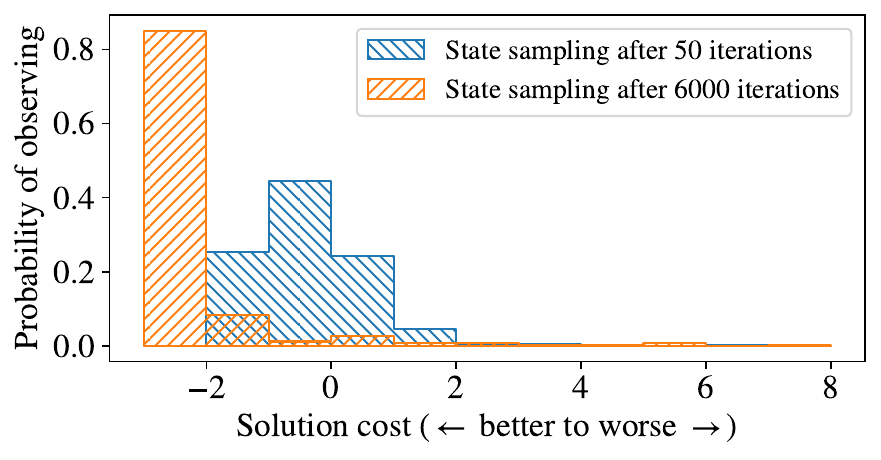}
    \label{fig:sampling}
\end{subfigure}
\caption{Numerical simulation of the thermalization of a system of 9 qubits under the MIS cost function for an Erd\H{o}s-R\'enyi graph with 0.7 edge probability. (a) The expectation value of the cost function is evaluated at each iteration for Markov chains with varying values of the disorder parameter $W$, and the acceptance rate (A.R.) is calculated. (b) A histogram of bitstring probability against cost found by sampling the state after 50 (blue) and 6000 (orange) iterations of the Markov chain, corresponding to the dotted vertical lines in (a). The disorder parameter is fixed to $W/J=200$. The characteristic exponential behavior of the sampled costs after 6000 iterations indicates the correct thermalization of the Markov chain.}
\label{fig:therm_samp}
\end{figure}

First, we analyze the thermalization behavior of the Markov chain and provide a numerical indication of how the acceptance rate (A.R.) can be effectively controlled by changing the disorder parameter $W$. Fig.~\ref{fig:thermalization} shows, for a system of 9 qubits, the expectation value of the MIS cost function as a function of the Markov chain length, for different values of the disorder parameter $W$, which acts as a bound to the random field in Eq.~\eqref{Eq.1}. Instances deep in the MBL phase -- that is, with large values of $W$ -- present a higher acceptance rate than those instances that lie closer to the thermalized phase. Comparing the lines $W/J=4$ (thermalized) and $W/J=400$ (very localized) in Fig.~\ref{fig:thermalization}, it is clear how the speed at which the Markov chain thermalizes is directly related to the acceptance rate. A very low acceptance rate allows big, but infrequent, jumps. Conversely, a high acceptance rate will result in many small moves, with the risk of reaching a regime of slow diffusive walk. The effectiveness of a Markov chain partially lies in the ability to tune the acceptance rate to ensure rapid convergence to the equilibrium distribution. These results show that such a level of control is easily achieved in our algorithm.

We note that the problem of local minima can affect the algorithm, as the Markov chain with $W/J=100$ in Fig.~\ref{fig:thermalization} appears to be stuck around a suboptimal local minimum.

As a further indication of proper thermalization of the Markov chain, Fig.~\ref{fig:sampling} shows the histogram of the quantum state sampled at iteration number 50 and 6000. The longer the chain, the better the system thermalizes around its equilibrium distribution, as indicated by the quantum state becoming exponentially concentrated around lower values of the cost function. The data shown is taken from the Markov chain of Fig.~\ref{fig:thermalization} with $W/J=200$.

\subsection{Solving optimization problems}

\begin{figure}[htpb]
\centering
\begin{subfigure}{.5\textwidth}
    \caption{}
    \includegraphics[width=\linewidth]{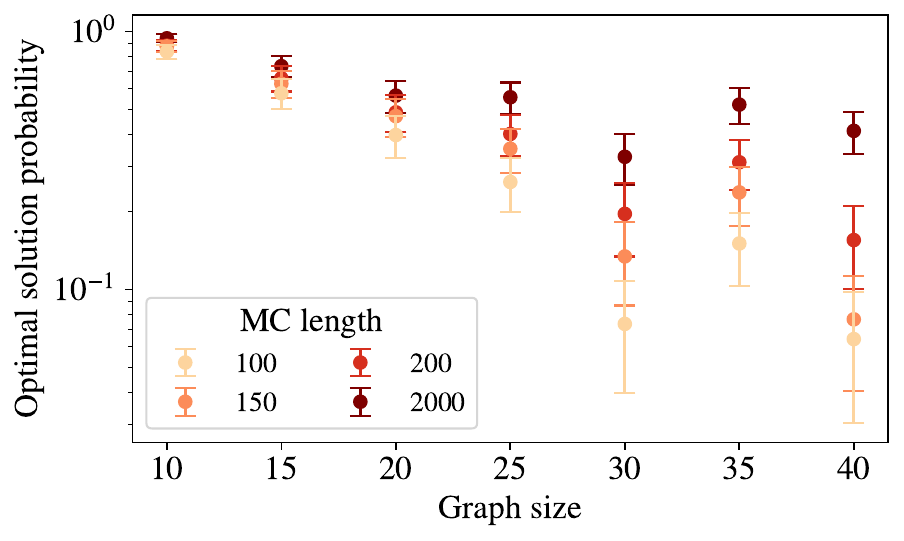}
    \label{fig:tts_mis}
\end{subfigure}
\begin{subfigure}{.5\textwidth}
    \caption{}
    \includegraphics[width=\linewidth]{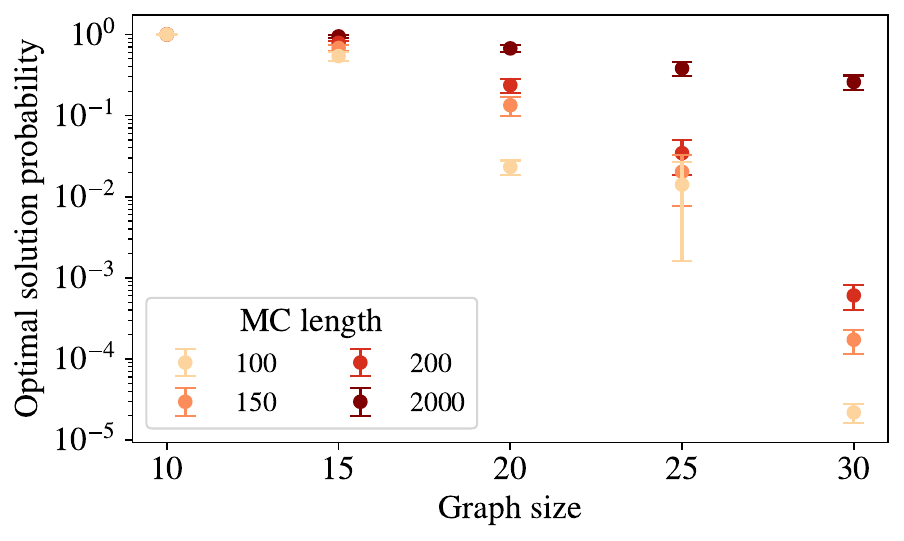}
    \label{fig:tts_maxcut}
\end{subfigure}
\caption{Probability of observing any of the optimal solutions to (a) MIS and (b) Max-Cut on Erd\H{o}s-R\'enyi graphs with 0.5 edge probability, assuming $10^4$ samples of the quantum state per iteration. Four different sets of points are shown, corresponding to stopping the Markov chain at length 100, 150, 200, and 2000. The acceptance rate for MIS ranges between 40\% (size 10) and 7\% (size 40). For Max-Cut, between 50\% (size 10) and 23\% (size 30).}
\label{fig:gtts}
\end{figure}

At low temperatures, the Gibbs distribution is dominated by the (often degenerate) ground state of the Hamiltonian, which represents the solution to the combinatorial optimization problem. Fig.~\ref{fig:gtts} shows the performance of the Markov chain in finding any of the optimal solutions to MIS (Fig.~\ref{fig:tts_mis}) and Max-Cut (Fig.~\ref{fig:tts_maxcut}) problems in a dataset of 35 random Erd\H{o}s-R\'enyi graphs with 0.5 edge probability of size varying from 10 to 40. The metric reported is the probability of observing at least once any of the solutions to the problem in the best quantum state explored during the algorithm, given a budget of $10^4$ samples of the quantum state taken at each iteration. We stress that this number is a conservative estimate of the probability of success, as it does not take into account all other slightly suboptimal quantum states traversed during the algorithm. Data are shown for Markov chains of variable length: 100, 150, 200, and 2000 iterations. In general, longer chains (more iterations) are required for larger graphs. Interestingly, even for chains of limited length, we can observe an optimal solution to the problem.

\begin{figure}[ht!]
\centering
    \includegraphics[width=\linewidth]{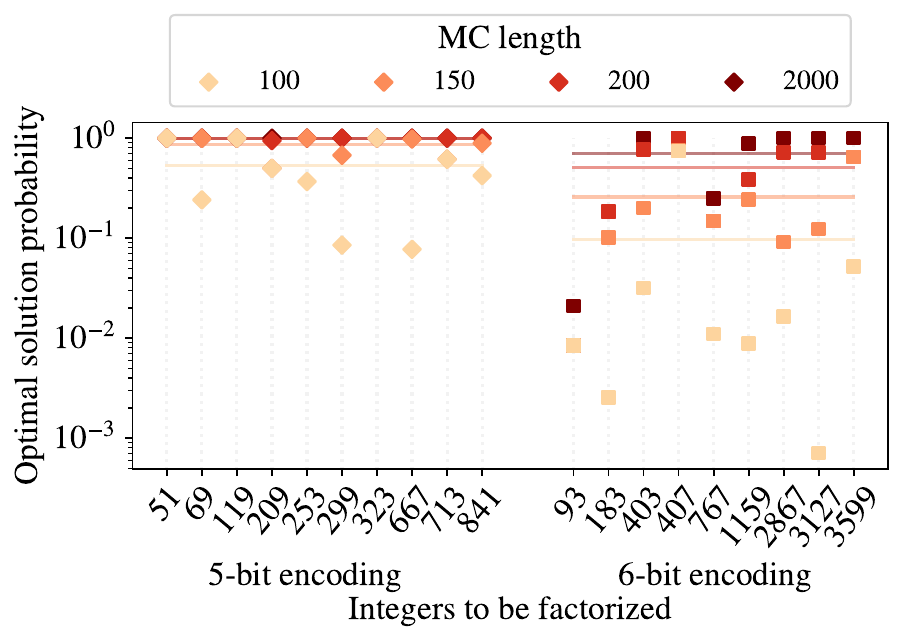}
    \caption{Probability of observing the optimal solution to the integer factorization problem. Each point on the x-axis represents an integer, $M$, for which an independent Markov chain is performed in order to factorize it. The prime factors are encoded with 5 bits for the group of integers on the left (diamond markers) and with 6 bits for the group of integers on the right (square markers). The plot shows the probability of observing the two correct prime factors for varying Markov chain lengths, assuming the state is measured $10^4$ times per iteration. The horizontal lines show the average probability for each Markov chain length. The acceptance rate is around 1\% for all instances.}
    \label{fig:factorization}
\end{figure}

Next, we turn our attention to a HUBO (Higher-order Unconstrained Binary Optimization) formulation of the prime factorization problem~\cite{factorization}, involving 4-body interaction terms (see Appendix \ref{sec: fact}). Solving a HUBO on a quantum annealer normally requires converting the higher-order terms into 2-body terms, resulting in an overhead in the number of qubits~\cite{hubo_qudratization, hubo_unfolding, hubo_tensor, hubo_set_cover, hubo_eig}. In our protocol, however, sampling from a general k-body Hamiltonian requires the same resources as sampling from a 2-body Hamiltonian and incurs no additional overhead. In Fig.~\ref{fig:factorization}, we show the results for the prime factorization of 19 random integers $M$ that are products of exactly two prime numbers $p$ and $q$. The size of the qubit register required is given by the number of bits used to encode the prime factors as binary numbers. If each prime factor is encoded in $n$ bits, the number of qubits is $2n$. Two sets of integers are factorized, corresponding to prime factors represented by 5 or 6 bits, and therefore, qubit registers of 10 and 12 qubits, respectively. The integer $M$ is chosen large enough so that it cannot be represented with $n$ bits to prevent the possibility that $1\cdot M$ is a solution. The algorithm is able to factorize all integers attempted with high probability (the longer the Markov chain, the higher the probability) with the exception of one of the 12 qubit instances ($M=93$), where the probability remained low throughout.

\section{Conclusion and outlook}

In this work, we introduced a novel quantum Markov chain algorithm that exploits the physics of many-body localization to facilitate efficient sampling of quantum states of qubit systems. We have shown how the choice of MBL is instrumental in ensuring that the necessary conditions for ergodicity are satisfied, and how the algorithm can be conveniently tuned to the desired acceptance rate by increasing of decreasing disorder.

The algorithm offers a way to extend the ability of analog quantum hardware  to solve sampling problems involving arbitrary qubit Hamiltonians, including Hamiltonians of order higher than quadratic. 

In the form presented in this paper, the algorithm only requires quantum hardware capable of simulating a 1D Ising chain with nearest-neighbour interactions. However, the need to evolve the initial quantum state through the whole chain at each iteration imposes stringent requirements on coherence times. 

We showcased the practical application of the method by applying it to well-known optimization problems: Maximum Independent Set, Max Cut, and Prime Factorization. A study of the scaling properties with system size was out of scope for the present investigation and will be the subject of future work.

We stress that the algorithm is not restricted to solving optimization problems. An immediate direction for future research would be to test its performance in producing Monte Carlo estimates of expectation values for Hamiltonians of interest, for use as a method for state preparation, or for sampling from a distribution of quantum states with desired properties.

Possible extensions of the method can involve different protocols for generating MBL unitaries, including the use of 2D systems. The faster spread of information could help reduce the chain length required for convergence to the equilibrium distribution.

Another potential direction is to train a small effective model on the samples of the current algorithm, for example a restricted Boltzmann machine or an Ising model~\cite{LiHuang2017, Liu2017}. Then, this model can be used to propose new states or serve as a good starting point for the chain. This can accelerate the convergence of the chain, reduce the number of measurements required, and also be combined with classical methods.

\clearpage 
\appendix


\section{Theory of general state space Markov chains and convergence bounds}\label{app:theory}

Markov chains are a class of algorithms that are used to sample from complex probability distributions by constructing a stochastic process that converges to the target distribution as its equilibrium state. In essence, it generates a sequence of samples in which each new sample depends only on the current state, enabling the exploration of high-dimensional spaces.

The Metropolis-Hastings algorithm is one of the most popular Markov chain methods. At each step, a new candidate state is proposed from a proposal distribution with density $q(x', x)$, where $x$ is the current state and $x'$ is the proposed state. The candidate is accepted with a probability:
$$
\alpha(x, x') = \min\left(1, \frac{\pi(x')q(x', x)}{\pi(x)q(x, x')}\right),
$$
where $ \pi (x) $ represents the target distribution. If the candidate is accepted, the chain moves to $x'$; otherwise, it remains at $x$. This acceptance rule ensures that the chain satisfies detailed balance, thereby guaranteeing convergence to the desired distribution.

In our work, the Metropolis-Hastings algorithm is employed in the Hilbert space of an $N$-qubit system, so that the samples $x$ are quantum states $\ket{\psi}$. The proposed samples are generated by quantum evolution of the current quantum state with MBL unitaries in a reversible manner, so that $q(\ket{\psi}, \ket{\psi}')=q(\ket{\psi}',\ket{\psi})$, and calculating the acceptance probability $\alpha(\ket{\psi}, \ket{\psi}')$ classically.

The space of quantum states of a $N$-qubit system, or the space of unit-length vectors, can be seen as the hypersphere $S^{2^{N+1}-1} \subset \mathbb{R}^{2^{N+1}}$ in the following way: decompose a generic quantum state on an orthonormal basis $\ket{\psi} = \sum_i c_i \ket{i}$, for $2^N$ complex amplitudes $c_i$ such that $\sum_i |c_i|^2=1$. Separate real and imaginary part $c_i = a_i + i b_i$, obtaining the normalization equation $\sum_i a_i^2 + b_i^2 = 1$, which is the equation of a sphere in $2^{N+1}$ real coordinates, hence $S^{2^{N+1}-1}$. Therefore, state space being a continuous space, one needs to use the theory of general state space Markov chains~\cite{nummelin}, rather than the more common theory of finite or countably infinite state spaces.

\subsection{Irreducibility}
For general state space Markov chains, irreducibility is taken with respect to a measure. More specifically, let $E$ be the state space of the Markov chain, equipped with a $\sigma$-algebra $\mathcal{E}$. For any $A\in \mathcal{E}$ and $x\in E$, define $P^n(x,A)$ to be the probability that, starting from $x$, the Markov chain reaches in $n$ iterations a state $y\in A$. Given a $\sigma$-finite measure $\varphi$ such that $\varphi(E)>0$, the Markov chain is called $\varphi$-irreducible if, for any initial state $x\in E$ and $A\in\mathcal{E}$ with $\varphi(A)>0$, one has $P^n(x,A)>0$ for some positive integer $n$~\cite{tierney}. Intuitively, $\varphi$-irreducibility demands that from any initial state one should be able to reach in $n$ steps any measurable region of state space.

We will prove $\phi$-irreducibility with respect to the Lebesgue measure $\mu_L$. Take $x\in E$ and $A\in\mathcal{E}$. For the Metropolis-Hastings kernel, call $q(x,y)$ the probability density of proposing a transition from $x$ to $y$, and $\alpha(x,y)$ the probability of accepting it. One has:
\begin{equation}
    P^n(x, A) \ge \int_A  \tilde{p}(x,y) \mu_L(dy)
\end{equation}
with
\begin{align}
    \tilde{p}(x,y):=\iint_E q(x,z_1) \alpha(x,z_1) q(z_1,z_2)\alpha(z_1,z_2) \ldots \notag \\
    \ldots q(z_{n-1}, y) \alpha(z_{n-1},y) \mu_L(dz_1)\ldots \mu_L(dz_{n-1})
\end{align}
where essentially the rejection part of the Metropolis kernel at each step was discarded because we are only interested in a lower bound. For equilibrium distributions of the type $\pi(\ket{\psi}) \sim \bra{\psi} O \ket{\psi}$ with $O$ a positive-definite operator as those discussed in the main text, one has that the acceptance probability is always strictly positive. A global minimum acceptance probability can be defined, which is strictly positive because state space $E$ is compact:
\begin{equation}
    \alpha_m := \frac{\inf_{x\in E} \pi(x)}{\sup_{x\in E}\pi(x)} > 0
\end{equation}
so that:
\begin{align}
    \tilde{p}&(x,y) \ge \notag \\ &\alpha_m^n \iint_E q(x,z_1) \ldots q(z_{n-1}, y) \mu_L(dz_1)\ldots \mu_L(dz_{n-1}).
\end{align}
The integral in $\tilde{p}(x,y)$ now contains only the proposal probabilities $q(\cdot,\cdot)$, and therefore it measures the probability of proposing a path of length $n$ from $x$ to $y$. Call this probability $q^{(n)}(x,y)$. Now we make use of two facts: 
\begin{enumerate}
    \item A sufficiently long random chain of MBL unitaries is CUE, i.e. the uniformly random measure on the unitary group;
    \item The unitary group $U(d)$ acts transitively on the hypersphere $S^{2d-1}$, i.e. for all $x, y\in S^{2d-1}$ it exists $u \in U(d)$ such that $u\cdot x = y$ (in our case $d=2^N$).
\end{enumerate}
Therefore, because of (2.) it always exists a unitary that brings from $x$ to $y$, and because of (1.), if $n$ is sufficiently large, there is a strictly positive probability of that unitary being the chain of proposed MBLs. Therefore for $n$ sufficiently large $q^{(n)}(x,y)>0$ and we can write:
\begin{align}\label{eq: irred}
    P^n(x, A) \ge \int_A  \tilde{p}&(x,y) \mu_L(dy) \ge \notag \\ &\alpha_m^n \int_A q^{(n)}(x,y) \mu_L(dy).
\end{align}
Now if we denote:
\begin{equation}
    q^{(n)}_m := \inf_{x,y\in E} q^{(n)}(x,y)
\end{equation}
we have $q^{(n)}_m >0$ because state space is compact, and therefore:
\begin{equation}\label{eq: irred}
    P^n(x, A) \ge \alpha_m^n q^{(n)}_m \int_A \mu_L(dy) = \alpha_m^n q^{(n)}_m \mu_L(A) > 0
\end{equation}
which proves $\mu_L$-irreducibility.

\subsection{Uniform ergodicity}
A Markov chain is uniformly ergodic if there is a positive constant $M$ and $0<r<1$ such that:
\begin{equation}\label{eq: uniferg}
    \sup_{x\in E}|| P^n(x,\cdot) - \pi|| \le Mr^n   
\end{equation}
meaning that convergence to the equilibrium distribution happens geometrically fast uniformly in $x$. A convenient way of verifying uniform ergodicity is to use the concept of small set. A subset $C\subset E$ is $(n_0,\epsilon,\nu)$-small if there exist $\epsilon>0$, a positive integer $n_0$, and a probability measure $\nu$ on $E$ such that the following minorization condition holds for all $x\in C$ and all measurable sets $A$:
\begin{equation}\label{eq: small set}
    P^{n_0}(x, A) \ge \epsilon \nu(A).
\end{equation}
A useful result is that if a Markov chain is such that the entire state space $E$ is $(n_0,\epsilon,\nu)$-small, then the chain is uniformly ergodic and $\sup_{x\in E}|| P^n(x,\cdot) - \pi|| \le (1-\epsilon)^{\lfloor n/n_0 \rfloor }$~\cite{tierney, nummelin, rob_ros}.

In the previous section on irreducibility, we have shown that, for $n_0$ large enough, $P^{n_0}(x, A) \ge \alpha_m^{n_0} q^{(n_0)}_m \mu_L(A)$ for any measurable $A$ and any $x\in E$, which automatically verifies the minorization condition \eqref{eq: small set} with $C=E$, meaning that state space is $(n_0, \epsilon, \mu_L)$-small, and the chain is therefore uniformly ergodic.

In order for the minorization condition to hold, $n_0$ needs to be large enough so that the chain of MBL unitaries approximates the CUE ensemble. In Appendix~\ref{app:cue} we show that the $n_0$ needed to verify the condition of CUE is $O(10)$, and decreases with increasing problem dimension. Therefore one can expect a $1-\epsilon$ factor improvement in convergence to the equilibrium distribution every few tens iterations of the Markov chain, with $\epsilon \ge \alpha^{n_0}_m q^{(n_0)}_m$.

\subsection{Bounds on mixing time}
Bounds on convergence can be formulated in terms of the acceptance probability $T(x)$:
\begin{equation}
    T(x) := \int_E q(x,y)\alpha(x,y)\mu_L(dy).
\end{equation}
For uniformly ergodic chains, Eq.~\eqref{eq: uniferg}, one has (\cite{brown} Lemma 3):
\begin{equation}
    r \ge 1-T(x).
\end{equation}
Moreover, defining the mixing time $\tau_\epsilon$ as the number of iterations required to reach a certain error threshold $\epsilon$:
\begin{equation}
    \tau_\epsilon := \inf_{n \in \mathbb{Z}^+} \left[ \sup_{x\in E} || P^n(x,\cdot) - \pi|| \le \epsilon  \right]
\end{equation}
one has~\cite{brown}:
\begin{equation}
    \tau_{\epsilon} \ge \ln(1/\epsilon) \left(\frac{1}{T(x)}-1 \right).
\end{equation}

A lower bound on the mixing time can also be formulated in terms of the spectral radius $\rho(P)$ of the transition kernel $P$ or, equivalently, its spectral gap $\Delta(P) := 1-\rho$. It was recently proven~\cite{woodard} for uniformly ergodic Markov chains that:
\begin{equation}
    \tau_\epsilon \ge \frac{\ln(2 \epsilon)}{\ln(\rho)}.
\end{equation}
For uniformly ergodic Markov chains one has $\rho<1$~\cite{rob_ros2}, and in particular the spectral gap is always finite. When the spectral radius $\rho \approx 1$, then $\ln(\rho)\approx \rho-1=-\Delta$ and the inequality can be expressed in terms of the inverse spectral gap instead:
\begin{equation}
    \tau_\epsilon \ge \ln\left(\frac{1}{2\epsilon}\right)\Delta^{-1}.
\end{equation}
Estimating $\rho$ or $\Delta$, and their functional dependence on system size, is an open problem in most cases. One way is to use Cheeger's inequality~\cite{low_sok}:
\begin{equation}
    1-k \le \rho \le 1-k^2/8
\end{equation}
where $k$ is the conductance:
\begin{equation}
    k := \inf_{A\in\mathcal{E}} \frac{\int_A P(x,A^c) \pi(dx)}{\pi(A)\pi(A^c)}
\end{equation}
which intuitively measures the worst bottleneck in transitioning from some subset $A$ of state space to its complement, normalized by their measure.
Another method for bounding $\rho$ involves computing a matrix approximation to the transition kernel~\cite{qin}. Exploring these possibilities is left as future work.

\section{Impact of finite measurements}
In the accept/reject step of the algorithm, one is required to calculate ratios of the type 
\begin{equation}\label{eq:acc_prob}
    \frac{\pi(\ket{\psi})}{\pi(\ket{\phi})}    
\end{equation}
where $\pi(\ket{\psi})\sim \bra{\psi}O\ket{\psi}=\langle O \rangle_\psi$ for a positive-definite observable $O$. In practice, those expectation values can only be calculated with limited accuracy due to the finite number of measurements that can be taken. If $\langle O \rangle_\psi$ is calculated up to additive error $\epsilon$, then it can be considered a normally distributed random variable with mean $\langle O \rangle_\psi$ and variance $\sim \epsilon^2$. The way the additive error propagates to the ratio can be quantified by looking at the distribution of the ratio. The distribution of the ratio of two normal random variables is extremely complicated in the general case~\cite{hinkley}, but in our case there exists a good normal approximation to it~\cite{diaz_rubio} with variance $\sim \epsilon^2$, which implies that the error on the ratio will also be of order $\epsilon$.

Now we want to provide an estimate of the number of measurements required to calculate the expectation values up to additive error $\epsilon$.

Let $K$ denote the number of measurements. Let $\langle O \rangle_\psi$ denote the true expectation value of $O$ on the state $\ket{\psi}$ and $\bar{O}_K$ the estimate of $\langle O \rangle_\psi$ obtained from $K$ measurements. One can use Chebyshev's inequality to calculate the probability that the error in estimating $\langle O \rangle_\psi$ with $\bar{O}_k$ is larger than a fixed threshold $\epsilon$:
\begin{equation}
    P\left(|\bar{O}_K - \langle O \rangle_\psi| \ge \epsilon \right) \le \frac{\sigma_O^2}{K \epsilon^2}
\end{equation}
where $\sigma_O^2 = \langle O^2 \rangle_\psi - \langle O \rangle_\psi^2$. If one requires this probability of failure to be smaller than a certain $\delta$, then it follows that it suffices to take:
\begin{equation}\label{eq:num_meas}
    K \ge \frac{\sigma_O^2}{\epsilon^2 \delta}.
\end{equation}
This scaling, although polynomial in $\epsilon$, might prove problematic in $\sigma_O^2$. For a concrete example, take $O$ to be $e^{-\beta H}$, with $H$ some Hamiltonian. The variance of $O$ can be bounded by:
\begin{equation}
    \sigma_O^2 \le \langle O^2 \rangle_\psi = \sum_b |\psi_b|^2 e^{-2\beta E_b}
\end{equation}
where $\ket{b}$ are eigenvectors of $H$ with eigenvalue $E_b$. If $E_m$ denotes the minimal energy of $H$, then the variance can be further bounded by:
\begin{equation}\label{eq:brutal_bound}
    \sigma_O^2 \le e^{-2\beta E_m}.
\end{equation}
In certain cases, when the ground state energy of $H$ scales logarithmically with system size, \eqref{eq:brutal_bound} and \eqref{eq:num_meas} imply that the number of measurements scales at worst polynomially in both $\epsilon$ and system size. However, in other cases of interest, the scaling can be exponential in system size. For instance, take $H$ to be the MIS cost Hamiltonian. Notice that the ground state energy of $H$ corresponds to minus the number of vertices in the MIS. As the MIS scales polynomially with the size of the graph~\cite{frieze, bollobas}, then it follows that the bound~\eqref{eq:brutal_bound} scales exponentially in the size of the problem. And therefore, because of \eqref{eq:num_meas}, the number of measurements might scale exponentially with system size.

However, the upper bound in Eq.~\eqref{eq:brutal_bound} might be too crude in practice. For instance, when the states sampled are close to the true ground state of $H$, and the amplitudes are exponentially concentrated on $E_m$ (see for instance Fig.~\ref{fig:sampling} and~\ref{fig:sampling_100}), then one might approximate, at first order in $\eta \sim 2^{-N}$:
\begin{align}
    \sigma_O^2 = \sum_b &|\psi_b|^2 e^{-2\beta E_b} - \left( \sum_b |\psi_b|^2 e^{-\beta E_b} \right)^2 \notag \\ 
    &\approx (1-\eta)e^{-2\beta E_m} + \eta C - \left((1-\eta)e^{-\beta E_m} + \eta C'\right)^2 \notag \\
    &\approx \eta e^{-\beta E_m} C''
\end{align}
suggesting that, once the Markov chain has thermalized, $\sigma_O^2 \sim O(1)$ and the number of measurements required is polynomial in $\epsilon$ and does not scale with system size.

\section{Ensemble statistics for products of MBL unitaries}\label{app:cue}
As highlighted in the main text, an important property of Markov chains is irreducibility, or the possibility of reaching the entire state space. We argue that in our algorithm, irreducibility is ensured by the fact that products of random MBL unitaries effectively belong the CUE (Circular Unitary Ensemble), i.e. the uniform measure in the space of unitary operators. Such measure is also called Haar random. Here, we present a numerical investigation of the speed of convergence of products of random MBL unitaries to CUE.

If $\{U_1, \ldots, U_M\}$ is a set of $M$ random unitary operators belonging to the MBL phase, we denote their product:
\begin{equation}
    \mathcal{U}_M := U_1 \cdot U_2 \cdot \ldots \cdot U_M   
\end{equation}
As an indicator of the phase that a certain ensemble of unitary operators belongs to, we use the level spacing statistics~\cite{level_spacing}. If $e^{i \theta_1}, \ldots, e^{i \theta_N}$ are the eigenvalues of a unitary operator ordered so that $\theta_1 < \theta_2 < \ldots < \theta_N$, define the gap $\delta_n := \theta_{n+1}-\theta_n$. The level spacing is defined as:
\begin{equation}
    r_n := \frac{\min [\delta_n, \delta_{n+1}]}{\max [\delta_n, \delta_{n+1}]} \in [0,1]
\end{equation}
The statistics can be collected over all values of $n=1,\ldots,N$, so that the spacing will be denoted simply $r$. The level spacing for unitaries in the MBL phase is expected to follow a Poissonian:
\begin{equation}
    Pr(r) \overset{\text{\tiny MBL}}{=} \frac{2}{(1+r)^2}
\end{equation}
while the level spacing for CUE can be calculated, in good approximation, by integrating the following expression (see Appendix B of~\cite{level_spacing} for details):
\begin{align}
    Pr(r) \overset{\text{\tiny CUE}}{=} \Theta(1-r)\frac{1}{\mathcal{N}} \int_0^{2 \pi} dx \int_0^{2\pi-x}dy \ \ \delta\left(r-\frac{x}{y}\right) \notag \\ \left( \sin \left( \frac{x}{2} \right) \sin \left( \frac{y}{2} \right) \sin \left( \frac{x+y}{2} \right) \right)^2
\end{align}
where $\mathcal{N}$ is a normalization constant, $\Theta$ is the Heaviside step function and $\delta$ is the Dirac delta.

To assess numerically the convergence of MBL unitaries to CUE, we calculate the full propagator from one period of Hamiltonians of the type of Eq.~\eqref{Eq.1}, using the same parameters used in the Results section: $B_0 = -\delta B = 1.25 J$, $\omega = 10J$, $W=200J$, $J = 4.15$.

\begin{figure}[htpb]
\centering
\begin{subfigure}{.5\textwidth}
    \caption{}
    \includegraphics[width=\linewidth]{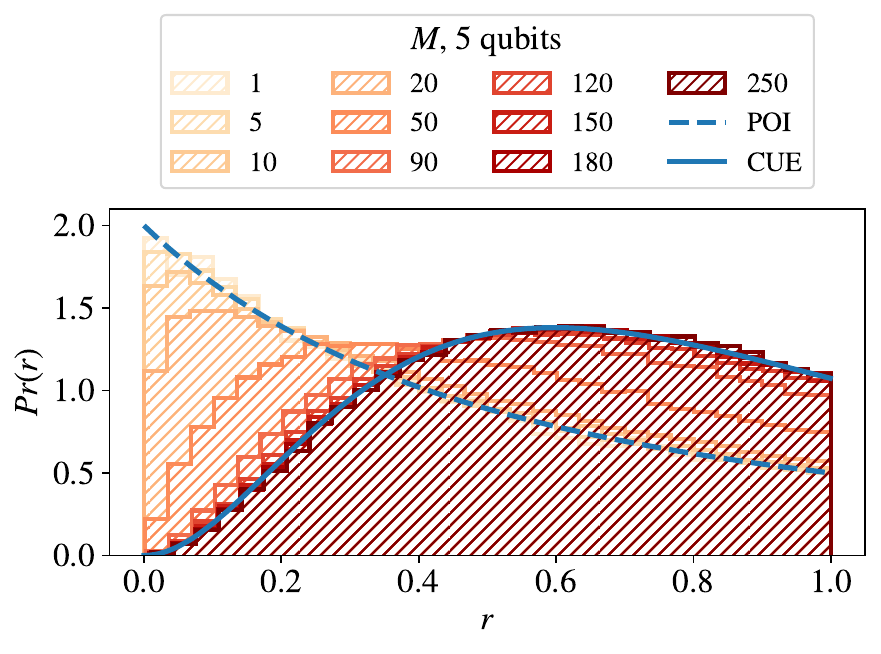}
    \label{fig:POI_CUE_hist}
\end{subfigure}
\begin{subfigure}{.5\textwidth}
    \caption{}
    \includegraphics[width=\linewidth]{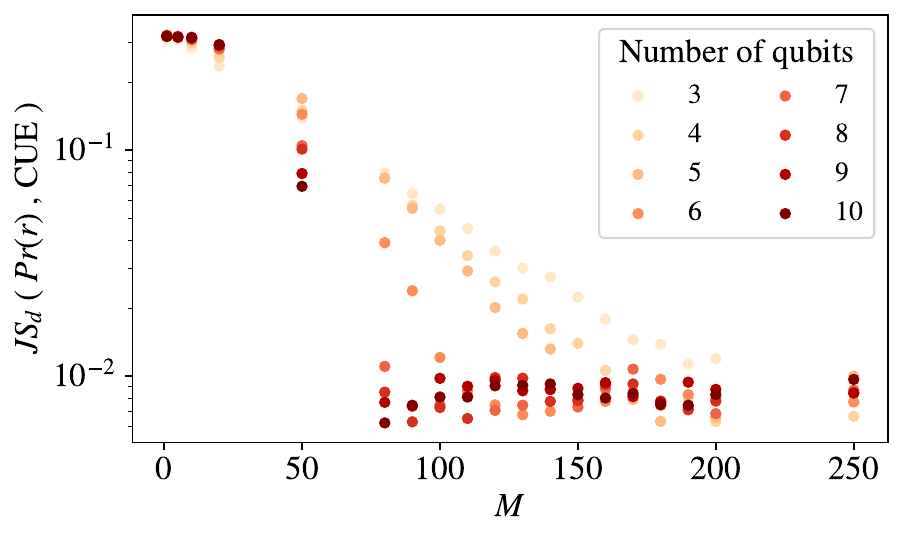}
    \label{fig:JS}
\end{subfigure}
\caption{Convergence of products of MBL unitaries to CUE statistics. The number of matrices in the product is denoted by $M$, and the resulting unitary is denoted $\mathcal{U}_M$. (a) For a system of 5 qubits, histograms of the level spacing of $\mathcal{U}_M$ are plotted for various values of $M=1,\ldots,250$. For $M=1$, the system shows Poissonian statistics typical of MBL (dashed blue line). As $M$ increases, the system quickly converges to CUE statistics (solid blue line). All histograms are generated by collecting the $2^5=32$ eigenvalues of $\mathcal{U}_M$, in an ensemble of 8000 propagators. (b) The Jensen-Shannon distance between the level spacing statistics $Pr(r)$ and CUE as a function of $M$, for systems of 3 to 10 qubits. The distance settles around a value of order $10^{-2}$ for all sizes, with larger systems converging more rapidly.}
\label{fig:CUE_convergence}
\end{figure}

In Fig.~\ref{fig:POI_CUE_hist} we show, for a system of 5 qubits, the convergence to CUE of the level spacing distribution of products of random MBL unitaries. The number of MBLs multiplied together, $M$, goes from 1 (where Poissonian statistics is expected) to 250. A good agreement with CUE statistics is reached approximately at $M=150$. Furthermore, we measure the agreement between $Pr(r)$ and CUE by calculating their Jensen-Shannon distance $JS_d$:
\begin{equation}
    JS_d(P,Q) := \sqrt{\frac{KL(P,A) + KL(Q,A)}{2}}
\end{equation}
where $P$ and $Q$ are distributions, $KL(\cdot,\cdot)$ is the Kullback-Leibler divergence, and $A$ is the point-wise average of $P$ and $Q$. One can see $JS_d$ as a metric obtained by symmetrizing the Kullback-Leibler divergence. In Fig.~\ref{fig:JS} we show how $JS_d(Pr(r), CUE)$ varies as a function of $M$ for different system sizes from 3 to 10 qubits. The speed of convergence increases with increasing system size, with only 90 MBLs needed for systems of 9 or 10 qubits to show to a good approximation CUE behavior.

The evidence reported here supports the claim of irreducibility of our Markov chain. It does not, however, provide an estimate of the mixing time of the Markov chain, since the algorithm is aimed at exploring the Hilbert space in a biased manner, as dictated by the Metropolis rule, and not just uniformly at random.

\section{Parameter selection}\label{app:parameters}
All tests reported in the main text are performed with a fixed set of parameters: $B_0 = -\delta B = 1.25 J$, $\omega = 10J$, and $J = 4.15$. The remaining parameter, the disorder strength $W$, was varied from $4J$ up to $400J$. Here we discuss more in depth how to choose these parameters, and their impact on the Markov chain.

\subsection{Disorder strength $W$}
We showed in the main text how $W$ is essentially the one parameter controlling the acceptance rate of the Markov chain, as it is inherently related to the transition between thermalization and localization. Therefore, all other parameters being fixed, $W$ should be chosen so that the acceptance rate of the Markov chain is the desired one. In the absence of specific reasons for choosing a particular value for the acceptance rate, a good starting point is around 23\%, a demonstrably optimal value~\cite{mixing_time_1, mixing_time_2, mixing_time_3} in a variety of situations, independently of system size. Parameter tuning can be done through preliminary runs or on the fly, for example by using dual averaging schemes as in~\cite{nuts}.

One question that might arise from Fig.~\ref{fig:thermalization} is why the chain with $W=100J$ is the one getting stuck in a local minimum, and if certain values of $W$ (or acceptance rate) are more susceptible than others to do so. The problem of local minima is a known one in Metropolis, and it is due to the diffusive walk nature of the algorithm. Although choosing a suitable acceptance rate definitely helps, it is really a shortcoming of the algorithm rather than parameter choice. More sophisticated strategies related to smart control of the inverse temperature $\beta$, such as simulated annealing or parallel tempering, can be employed within the Metropolis framework to alleviate the problem. Such improvements are fully compatible with our algorithm, but they were not pursued here.

\begin{figure}[htpb]
\centering
\begin{subfigure}{.5\textwidth}
    \caption{}
    \includegraphics[width=\linewidth]{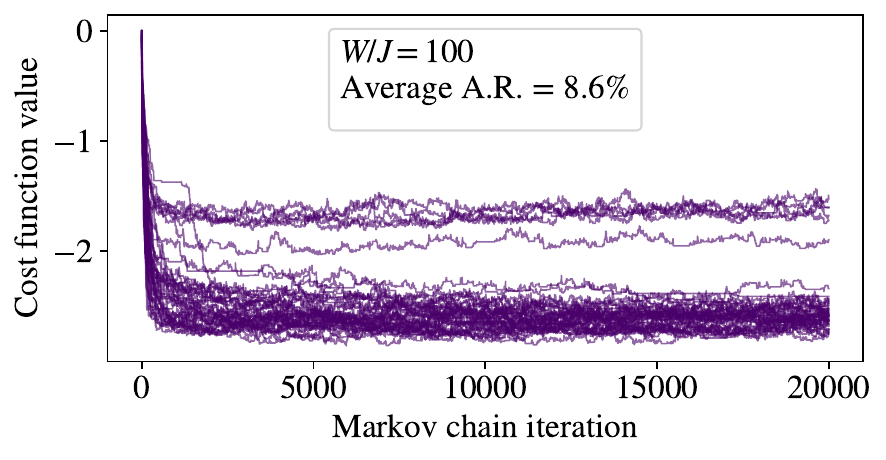}
    \label{fig:therm_100}
\end{subfigure}
\begin{subfigure}{.5\textwidth}
    \caption{}
    \includegraphics[width=\linewidth]{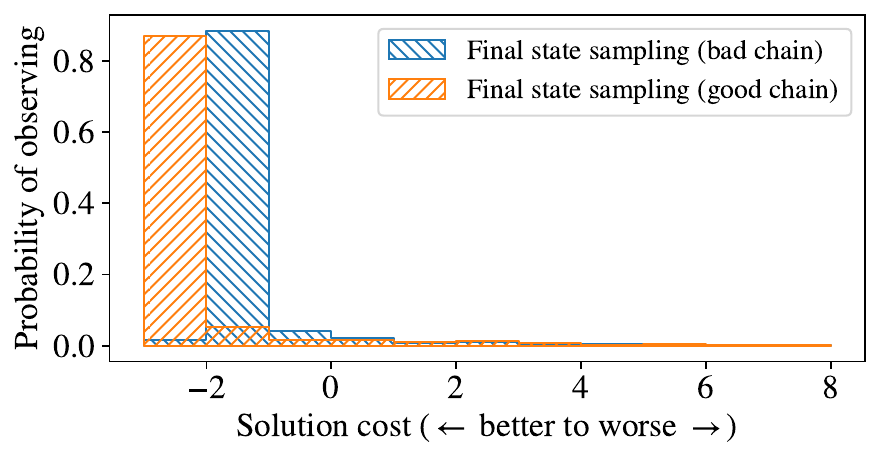}
    \label{fig:sampling_100}
\end{subfigure}
\caption{Thermalization of a system of 9 qubits under the MIS cost function for an Erd\H{o}s-R\'enyi graph with 0.7 edge probability (same as Fig.~\ref{fig:therm_samp}). (a) Thermalization plot for a collection of 40 independent Markov chains with $W=100J$. The plot shows how most chains correctly thermalize around the true minimum, while some other remain stuck around a local minimum. (b) Final state sampling for a chain that has thermalized correctly (orange) and a chain that remained stuck around a local minimum (blue). The quantum states sampled by the suboptimal chain are peaked around MIS-1, the next-best solution to the MIS problem.}
\label{fig:multiple_100}
\end{figure}

To show that indeed $W=100J$ is not particularly problematic as a value, we show in Fig.~\ref{fig:therm_100} 40 independent Markov chains with $W=100J$ under the MIS cost function. Most chains correctly thermalize to the correct minimum, while some others are stuck in a suboptimal configuration. To give a visual interpretation of what the suboptimal configuration is, we show in Fig.~\ref{fig:sampling_100} the final state sampling of two chains, one that has thermalized correctly and one that has not. The correctly thermalized chain ends up sampling quantum states exponentially peaked around the correct value of the MIS. The incorrectly thermalized chain, on the other hand, has settled around a configuration representing MIS-1, the next-best solution to the MIS problem.

\subsection{Drive frequency $\omega$ and nearest-neighbour interaction $J$}
On paper, the one important parameter of Metropolis is the acceptance rate, and therefore one could argue that $\omega$ and $J$ can be set almost arbitrarily, as long as a suitable value for $W$ can then be found to bring the acceptance rate to the desired value. However, the choice of $\omega$ and $J$ holds an important practical value.

Every quantity is expressed in units of $J$, and therefore it is important in implementations on real hardware that the nearest-neighbour coupling is set to a value that would result in realistic field strengths for the $X$ and $Z$ fields of the Hamiltonian.

The drive frequency, $\omega$, plays a practical role in determining the length $T$ of each single quantum evolution block through the relation $T = 2\pi/\omega$. A Markov chain of length $M$ will therefore translate to a total time evolution of length $M\times T$, determining the coherence time required for the hardware to carry out the whole chain.

As an example, take the parameters used in the main text, $B_0 = -\delta B = 1.25 J$, $\omega = 10J$, $W=200J$, and $J = 1.04$. Implementing those on a neutral atom QPU first requires mapping the Hamiltonian~\eqref{Eq.1} to the native Ising Hamiltonian of neutral atoms~\cite{neutralatoms}:
\begin{equation}\label{eq: na_ham}
    H_{ryd} = -\sum_{i=1}^N \delta_i n_i + \frac{\Omega(t)}{2}\sum_{i=1}^N X_i + \frac{C_6}{a^6} \sum_{i=1}^{N-1} n_i n_{i+1}
\end{equation}
where $n_i = (Z_i + 1) / 2$, $C_6$ is a physical constant that depends on the Rydberg level used for the excited state, and $a$ is the spacing between atoms. In Eq.~\eqref{eq: na_ham} we have neglected all couplings beyond nearest-neighbor, which are suppressed by a factor of at least $2^6$. A little algebra gives:
\begin{align}
    J &= \frac{C_6}{4 a^6} \\
    B(t) &= \frac{\Omega(t)}{2} \\
    \text{for }i=2,\ldots,N-1 \quad \delta_i &= 4J - 2 h_i\\
    \text{for }i=1,N \quad \delta_i &= 2J - 2 h_i
\end{align}
In neutral atom terminology, $\delta_i$ is called detuning. Notice how the detuning in the bulk is different from the detuning on the border. One could obtain $J=1.04 \ rad/\mu s$ by setting $a=7.7 \mu m$ and Rydberg level $60$. This would then translate to a longitudinal field $\Omega(t)$ oscillating between 0 (when the amplitude is $2B_0-2\delta B$) and $5.2 \ rad/\mu s$ (when the amplitude becomes $2B_0+2\delta B$), which is well within what is technologically possible. As for the detuning, it would be a number between -203 and +211 $rad/\mu s$, slightly suboptimal since current hardware supports detunings in a range of around $\pm$150 $rad/\mu s$. The oscillation frequency controls the duration of each MBL block, and $\omega=10J$ would translate to a quantum evolution of $T = 2 \pi / \omega = 0.6 \mu s$, which is fine for $O(10)$ iterations, but would not allow a meaningfully long Markov chain since the current experimental limit for the whole evolution is around $6 \mu s$. To bring the numbers down to a range of experimental feasibility, one could increase the interatomic spacing (to lower $J$ and hence $W$), and increase the frequency $\omega$ (to lower $T$). Achievable numbers would be $a=8$ and $\omega\sim 100J$, which would allow $O(100)$ iterations. However, note that the drive frequency on physical hardware is inherently limited. In neutral atom architectures, for example, the frequency is limited by the risetime of the acousto-optic modulator (AOM). $O(100)$ iterations in $6\mu s$ is within current technological capabilities, but close to the upper limit.

\subsection{Drive amplitude $\delta B$ and $B_0$}

When tuning parameters like $J$ and $\omega$ for experimental purposes, one should always keep in mind the thermalization-localization phase diagram, as shown in Fig. 2 of~\cite{PhysRevB.103.165132}, in order to make sure that the parameters result in an MBL unitary. The diagram depends only on the dimensionless quantities $\omega/J$ and $W/J$, which leaves the choice of $J$ free, but it potentially depends on the drive amplitude $\delta B$ and $B_0$.

For the simulations presented in the main text, we chose $B_0=-\delta B=1.25J$ as a matter of convenience. The fact that the two values are equal in absolute value makes it so that the drive stays positive, oscillating between 0 and $2B_0$. The specific value $1.25J$ was chosen because it matches the one used in~\cite{PhysRevB.103.165132}, which allowed us to reuse the phase diagram computed there. In this section we present a short discussion on the potential impact of using a stronger drive $B_0=-\delta B=5.25J$.

\begin{figure}[htpb]
\centering
\begin{subfigure}{.5\textwidth}
    \caption{}
    \includegraphics[width=\linewidth]{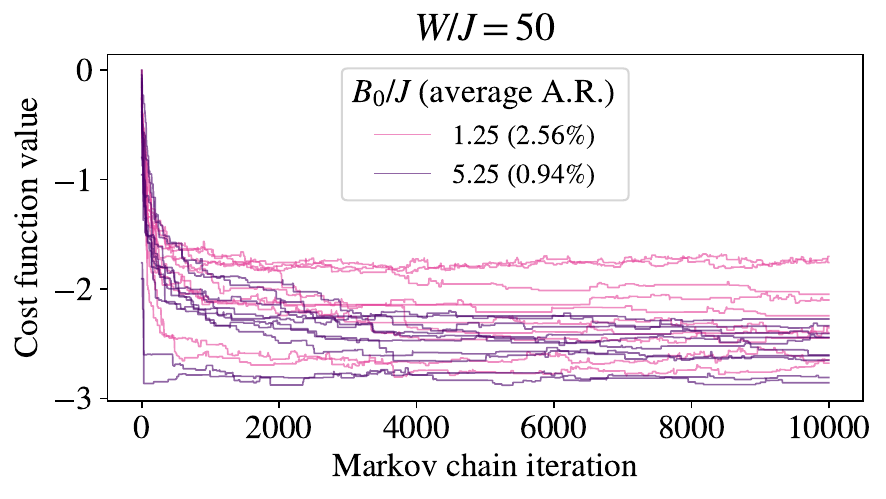}
    \label{fig:therm_50_comp}
\end{subfigure}
\begin{subfigure}{.5\textwidth}
    \caption{}
    \includegraphics[width=\linewidth]{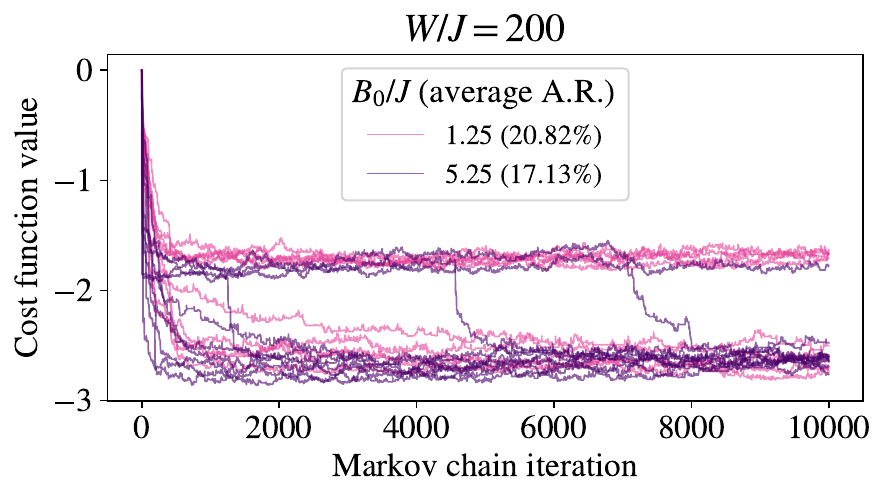}
    \label{fig:therm_200_comp}
\end{subfigure}
\caption{Effect of driving amplitude on the thermalization of a system of 9 qubits under the MIS cost function for an Erd\H{o}s-R\'enyi graph with 0.7 edge probability (same as Fig.~\ref{fig:therm_samp}). Ten independent Markov chains are shown for each value of driving strength: $B_0=-\delta B=1.25$ (pink) and $B_0=-\delta B=5.25$ (purple). Two disorder strengths are shown, (a) $W=50J$ and (b) $W=200J$.}
\label{fig:therm_comp}
\end{figure}

In Fig.~\ref{fig:therm_comp} we show an ensemble of Markov chains for two values of disorder strength, $W=50J$ and $200J$, and two values of drive amplitude $B_0=-\delta B =1.25J$ and $5.25J$. The two amplitudes will be referred to as weak and strong respectively. We observe how the acceptance rate is on average lower for the strong drive. However, similar simulations (not shown here) suggest that this trend reverses somewhere between $W=200J$ and $400J$. The strong drive seems to produce a few tunneling events in the $W=200J$ plot, where the chain manages to escape from the suboptimal minimum to the true one. One could argue that a stronger transversal field might indeed induce more frequent transitions, increasing the chance of tunneling. This phenomenon could be of potential interest for the method, providing a way to escape local minima. However, the statistics collected here is not enough to draw definite conclusions. Further studies are needed to confirm this, but they are outside the scope of the present paper.

\section{Effect of noise}
Since the protocol can be implemented on NISQ analog hardware, in this section we briefly discuss the effect of typical noise sources in a neutral atom implementation.

Referring to the physical Hamiltonian \eqref{eq: na_ham}, we distinguish the following main noise sources~\cite{noise, noise_in_prep}:
\begin{enumerate}
    \item Laser intensity, frequency, and phase fluctuations affecting $\Omega(t)$ and $\delta_i$
    \item Atomic position errors affecting $a$
    \item Coupling with the environment
    \item SPAM errors
\end{enumerate}

Laser fluctuations are in general of two types: high frequency (affecting the intensity and detuning within a single shot) and low frequency (causing shot-to-shot variations in intensity and detuning). High-precision numerical simulations~\cite{noise_in_prep} show how the decoherence induced by such noise sources is somewhat suppressed by operating in a regime of weak interaction, where the interatomic spacing is of the order $10 \mu m$.

Shot-to-shot variations in atomic positions affect the interatomic spacing $a$. In the worst case scenario, atoms might present an off-plane displacement of as much as $3\mu m$ from their expected position~\cite{noise_in_prep}. Again, the most effective way to mitigate this source of uncertainty is to increase the spacing $a$.

Coupling of the system with the environment translates in effective noise channels causing decay from $\ket{1}$ to $\ket{0}$ and dephasing. Such contributions start becoming important at timescales of about $10 \mu s$.

Preparation errors involve atom loss in the register building process. Typically, such errors can be corrected by post-selection, filtering out the faulty samples. Measurement errors include false positive and false negative detection errors. These are common to most quantum computing architectures, and consequently several mitigation techniques have been developed such as confusion matrix inversion.

\section{Cost functions}
\subsection{MIS}\label{sec: mis}
Let $G=(V,E)$ be a graph of size $N$ defined by its vertex set $V=\{1,\ldots,N\}$ and edge set $E\subseteq V\times V$. Let $\vec{x}=(x_1, \ldots, x_N)\in \{0,1\}^N$ be a binary vector of length $N$.

A solution to the Maximum Independent Set problem is given by any vector $\vec{x}^*$ that minimizes the following quadratic cost function:
\begin{align}
    \vec{x}^* = \argmin_{\vec{x}\in \{0,1\}^N} \left( -\sum_{i\in V} x_i + 2\sum_{(i,j)\in E} x_i x_j \right).
\end{align}
Given a solution $\vec{x}^*$, the corresponding independent set is $\{i\in V: x_i^*=1\}$.

\subsection{MAX-CUT}\label{sec: maxcut}
Let $G=(V,E)$ be a graph as above.

A solution to the Maximum Cut problem is given by any vector $\vec{x}^*$ that minimizes:
\begin{align}
    \vec{x}^* = \argmin_{\vec{x}\in \{0,1\}^N} \left( -\sum_{i\in V} N_i x_i + 2\sum_{(i,j)\in E} x_i x_j \right).
\end{align}
Where $N_i=\left|\{j\in V: (i,j)\in E\}\right|$ is the degree (number of neighbours) of vertex $i$.

\subsection{Prime Factorization}\label{sec: fact}
This derivation follows~\cite{factorization}.

Let $N$ be an integer that we wish to factorize into two primes $p$ and $q$. The prime factorization problem can be formulated as a least-squares problem by minimizing the difference between the product $pq$ and $N$:
\begin{equation}
\argmin_{p,q\in \mathbb{N}} (pq - N)^2.
\end{equation}
The minimum of this expression is 0, and it is achieved when $pq = N$.

Expanding the square yields:
\begin{equation}
\argmin_{p,q\in \mathbb{N}}(pq - N)^2 = \argmin_{p,q\in \mathbb{N}}(p^2q^2 - 2Npq + N^2).
\end{equation}
The $N^2$ term can be neglected since it is constant with respect to $p$ and $q$, giving the following equivalent cost function:
\begin{equation}\label{eq:factcf}
\argmin_{p,q\in \mathbb{N}}(p^2q^2 - 2Npq).
\end{equation}
This expression is quadratic in $p$ and $q$. However, in order to map this function to a binary minimization problem, we represent $p$ and $q$ in binary (radix-2) form:
\begin{equation}
p = \sum_{l=0}^{n-1} 2^l\, q_l, \quad q = \sum_{l=0}^{n-1} 2^l\, q_{n+l},
\end{equation}
where each $q_i \in \{0,1\}$. Substituting these expressions into the cost function~\eqref{eq:factcf} yields a polynomial in the binary variables. This polynomial typically includes quadratic, cubic, and quartic terms, thereby defining a Higher-Order Unconstrained Binary Optimization (HUBO) model.

\bibliography{references}
\bibliographystyle{apsrev4-1}

\end{document}